\documentstyle[12pt,amsmath]{article}
\begin{document}

\title{Fractional Quantum Hall Effect 
and vortex lattices.}

\author{S. V. Iordanski}

\date{Landau Institute for Theoretical Physics, 
Russian Academy of Sciences,Kosygin str.2 Moscow, 117334 Russia}

\maketitle

\begin{abstract}

It is demonstrated that all observed fractions at moderate Landau level fillings 
for the quantum Hall effect can be obtained without recourse to the
phenomenological concept of composite fermions. The possibility to have
the special topologically nontrivial many-electron wave functions is
considered. Their group classification indicates the special values of
of electron density in the ground states separated by
a gap from excited states. 

\end{abstract}

73.43.-f

\vspace{0.5cm}

The experimental discovery of Integer Quantum Hall Effect (IQHE) by
K.v Klitzing (1980) and Fractional Quantum Hall Effect (FQHE) by Tsui, Stormer
and Gossard (1982) was one of the most outstanding achievements in condensed
matter physics of the last century. 

Despite the fact that more than twenty 
years have elapsed since the
experimental discovery of quantum Hall Effect (QHE), the theory of
this phenomenon is far from being complete (see reviews [1, 2]).
This is primarily true for the Fractional Quantum Hall Effect (FQHE),
which necessitates the electron--electron interaction and can by no
means be explained by the one-particle theory, in contrast to the
IQHE. The most successful variational many-electron wave
function for explaining the 1/3 and other odd inverse fillings was constructed 
by Laughlin[3, 4]. The explanation of other observed fractions was obtained by
various phenomenological hierarchial schemes with construction of the "daughter"
states from the basic ones (Haldane 1983,Laughlin 1984, B.Halperin 1984).

 In those works, the approximation of extremely high magnetic
field was used and all states were constructed from the states at
the lowest Landau level. However, this does not conform to the
experimental situation, where the cyclotron energy is of the order of
the mean energy of electron--electron interaction. Moreover, this
approach encounters difficulties in generalizing to the other
fractions. Computer simulations also give a rather crude approximation 
for the realistic multiparticle functions, because the
number of particles in the corresponding calculations on modern
computers does not exceed several tens.

The most successful phenomenological description is given by the
Jain's model of "composite" fermions [5, 6], which predicts the
majority of observed fractions. According to this model, electrons
are dressed by magnetic-flux quanta with magnetic field 
concentrated in an infinitely narrow region around each electron. It
is assumed that even number of flux quanta provides that these
particles are fermions. The inclusion of this additional
magnetic field in the formalized theory leads to the so-called Chern--
Simons Hamiltonian. This approach is described in details in [7].

However, this theory gives an artificial 6-fermionic interaction whereas 
the actulal calculations use quite crude mean field approximation of the 
"effective" magnetic field as the sum of the external magnetic field and some 
additional artificial  field that provides the total magnetic flux quanta
 in accordance with Jain's model of composite fermions.

In the present work I shall show  how to remove some restrictions of 
 Jain-Chern-Simons model and obtain a more 
general and more simple model which does not change the  standard Coulomb 
interaction of electrons. The main
concept is associated with the notion of topological classification of
 quantum states. There is a number of
topological textures in condensed matter physics: Vortex lattices in a rotating 
superfluid, Abrikosov vortices in
superconductors, skyrmions in 2d electron systems at integer fillings of 
Landau levels. It is difficult to give an exact 
topological classification of the multiparticle wave function for various 
physical systems. Possibly the most simple
and general definition can be done using canonical transformation of 
the field operators of the second quantization.
 The canonical transformation of the field operators is one which does not 
 change their commutation relations.
 I do not consider the statistical transmutations which possibly can 
not be achieved at low energies 
considered in condensed matter physics. In general there must be the proper 
topological classification of  the canonical transformations itself.

In this work I consider the simplest case of the fermion canonical
 transformation not including spin degrees
of 
freedom and assuming the full polarization of 2d electrons
$$\psi({\bf r})=e^{i\alpha({\bf r})}\chi, \psi^{+}({\bf r})=\chi^{+} 
e^{-i\alpha({\bf r})}$$
with  $\alpha({\bf r}$ ) having vortex kind singularities. It is evident that 
 $\chi$ and $\chi^{+}$ satisfy Fermi kind 
commutation relations if $\psi$ and $\psi^{+}$ satisfy them. Inserting these
 expressions into the  standard
hamiltonian for the interacting  electrons (with omited spin indices )

\begin{multline}
\label{ham}
H=\frac{\hbar^2}{2m}\int \psi^{+}(-i{\bf \nabla }-\frac{e}{c\hbar}{\bf A})^2
\psi d^2r+\\
\int \frac{U({\bf r-r'})}{2}\psi^{+}({\bf r})\psi^{+}({\bf r'})
\psi({\bf r'})\psi({\bf r})d^2rd^2r'
\end{multline}
we get  a new Hamiltonian

\begin{multline}
\label{nham}
H=\frac{\hbar^2}{2m}\int \chi^{+}(-i{\bf \nabla}+ {\bf \nabla}\alpha -
\frac{e}{c\hbar}{\bf A})^2\chi d^2r\\
+\int \frac{U(|{\bf r-r'}|)}{2}\chi^{+}({\bf r})\chi^{+}({\bf r'})
\chi({\bf r'})\chi({\bf r})d^2rd^2r'
\end{multline}
where $U(r)$ is Coulomb interaction. I want to consider a set of periodic  
vortexlike singularities in
${\bf \nabla}\alpha$. Vector ${\bf \nabla}\alpha$ can be expressel in terms of
 Weierstrass zeta function
 used in
the theory of the rotating superfluids \cite{tk} given by the converging series
\begin{equation}
\label{z}
\zeta=\frac{1}{z}+\sum_{T_{nn'}\ne 0}(\frac{1}{z-T_{nn'}}+\frac{1}{T_{nn'}}+
\frac{z}{T_{nn'}^2})
\end{equation}
where $z=x+iy$ is a complex coordinate on 2d plain, $T_{nn'}=n\tau+n'\tau'$ 
and $\tau,\tau'$ are  the minimal
complex periods \cite{ui} of the vortex lattice. The phase factor $e^{i\alpha
}$ will be simple function on
2d plain if ${\bf \nabla}\alpha=K(Re\zeta,Im\zeta)$ and
\begin{equation}
\label{al}
\alpha({\bf r})=K
\int_{\bf r_0}^{\bf r}(Re\zeta dx+Im\zeta dy)
\end{equation}
with integer $K$ of any sign. The quantity $K$ and the periods $\tau,\tau'$
define the topological class of multiparticle
wave function. The transformed Hamiltonian (\ref{nham}) can not be restored to 
the ininitial form (\ref{ham}) 
by any smooth finite transformation of the function $\alpha$ .
That
 makes it topologically stable.
I shall investigate the pecularities of the ground state and excitations for 
this model at low temperature.

 Having in mind large magnetic fields it is interesting to consider the 
 simplified version of the hamiltonian
 (\ref{nham}) without the interaction term
 
 \begin{equation}
 \label{h'}
 H'=\frac{\hbar^2}{2m}\int\chi^{+}[-i{\bf \nabla}+{\bf \nabla}\alpha-
 \frac{e}{c\hbar}{\bf A(r)}]^2\chi d^2r
 \end{equation}
 
 This Hamiltonian has  properties very close to the Hamiltonian with a 
 constant magnetic field. Indeed the translation on any period ${\vec \tau}$ of 
 the 
 vortex lattice gives an additional constant in the brackets
 \begin{multline}
 \label{tr}
 {\bf r}\to{\bf r}+{\vec \tau}\\
 [-i{\bf \nabla}+{\bf \nabla}\alpha-\frac{e}{c\hbar}A({\bf r})]\to\\
 [-i{\bf \nabla}+{\bf \nabla}\alpha-\frac{e}{c\hbar}{\bf A}({\bf r})+
 {\vec \delta}({\vec \tau})-\frac{e}{c \hbar}{\bf A}(\vec \tau)]
 \end{multline}
 due to the properties of Weierstrass function 
 $\zeta(z+ \tau)=\zeta(z)+\delta(\tau)$ and the linear dependence of the 
 external vector potential ${\bf A}({\bf r})$ at constant magnetic field. The
 additonal constant terms can be removed by the gauge transformation of the
 field operators $\chi,\chi^{+}$. Thus the proper magnetic translation does
 not change Hamiltonian (\ref{h'}).
 
 If we introduce the "effective" vector potential 
 ${\bf A}_{eff}={\bf A}-\frac{c\hbar}{e}{\bf  \nabla}\alpha$, the magnetic 
 translation is
  given
 by the transformation
 \begin{equation}
 \label{mt}
 T_m({\vec \tau})\chi=\chi({\bf r}+{\vec\tau}) \exp(\frac {ie}{c\hbar}
 {\bf A}_{eff}({\vec \tau}){\bf r})
 \end{equation}
 for any real period of the vortex lattice.
 
  It is easy to connect ${\bf A}_{eff}({\vec \tau})$ with the "effective" 
  magnetic flux through the unit cell of the vortex lattice given by the 
  contour
  along it's boundaries
  $$\Phi=\oint{\bf A}_{eff}d{\bf r}= {\bf A}_{eff}({\vec \tau_1})
  {\vec \tau_2}-
  {\bf A}_{eff}({\vec \tau_2}){\vec \tau_1}$$
  On the other hand it can be calculated directly using the definition of
  ${\bf A}_{eff}$
  \begin{equation}
  \label{fi} 
  \Phi={\bf B_0}{\vec \tau}_1\times {\vec \tau}_2+K\Phi_0
  \end{equation}
  where $\Phi_0=2\pi\frac{e}{c\hbar}$ is the quantum of the flux,$B_o$ is
  the external magnetic field.
  
  As was shown by E.Brown (1964) \cite{br}, J.Zak (1964) \cite{zk}(see also
  \cite{lp9})
  
  the simple finite representation of the ray group of magnetic translations
  can be obtained only for rational number of the flux quanta per unit cell
  \begin{equation}
  \label{phi'}
  \Phi=\frac{l}{N}\Phi_0=B_0s+K\Phi_0
  \end{equation}
  where $s$ is the area of the unit cell of the vortex lattice, $l$ and $N$
  are integers without common factors.
  
  Thus  the situation for the vortex lattices  is isomorphous to 
  the case of uniform magnetic field with  a rational number of the flux quanta
  per the unit cell. Therefore it is possible to use all the argumentation
  following the paper \cite{br} in constructing of the finite representation
  for the ray group of magnetic translations. In order to construct the finite
  representation one must impose certain boundary conditions on the solutions 
  of Schroedinger equation with the hamiltonian (\ref{h'}). The simplest is
  the magnetic periodicity
  \begin{equation}
  \label{mp}
  T_m({\bf L})\chi({\bf r})=\chi({\bf r})
  \end{equation}
  where ${\bf L=L_1,L_2}$ define the size of the sample,
   ${\bf L_1}=NM_1{\vec \tau_1}$,${\bf L_2}=NM_2{\vec \tau_2}$ with integer
   $M_1,M_2$. It easy to show that any magnetically translated   function
   $\chi$ according to (\ref{mt}) will also satisfy (\ref{mp}). The simplest
   realization is the vortex lattice  consisting of exactly $N\times N$ unit
   cells.
   
   This conditions is the analog of Born-von Karman conditions in the absence
   of magnetic field. Indeed in a large enough system the density of states 
   practically does not depend on the exact form of boundary conditions. But 
   the restriction to the finite representations is important.
   
   The matrices of the representation are
   \begin{multline}
   \label{rp1}
   D_{jk}(0)=\delta_{jk}\\
   D_{jk}({\vec \tau_1})=\delta_{jk}\exp{i(j-1)\frac{l}{N}}\\
   D_{j,k}({\vec \tau_2})=\delta_{j,k-1}  \\
   (mod  N)(j,k=1,2...N)
   \end{multline}
   and the general matrix of the representation
   \begin{equation}
   \label{rp2}
   D_{jk}(n_1{\vec \tau_1}+n_2{\vec \tau_2})=\exp{i\pi\frac{ln_1}{N}[n_2+2(j-1)]}
   \delta_{j,k-n_2}\\(mod  N)
   \end{equation}
   
   The traces of all matrices are zero except identity which has a trace equak
   to
   $N$. The sum of the squares of traces is $N^2$ . Therefore the representation
   is irreducible . The square of the dimensionality is also $N^2$ therefore 
   there can be no other nonequivalent representation. The dimensionality of
   the representation  gives  $N$ fold degeneracy of the energy levels for
   Hamiltonian \ref{h'}. The number of the equivalent representations in a
   regular representation is also $N$. These equivalent representations 
   correspond to the states with different energies  for the real "crystal"
   containing not only  vortices but also a periodic potential. But for the
   simplyfied Hamiltonian (\ref{h'}) all $N^2$ translations are on equal footing
   and do not change the energy of the state 
   because they commute with the hamiltonian but do not commute with the each 
   other. Therefore all $N^2$
   elements of the regular representation must have the same energy.
   
   If $|l|\neq 1$ there is a possibility to have $l$ different periodic
   solutions, corresponding to the different number of the zeros for the wave
   function
   inside the magnetic cell having $l$  flux quanta, that is analogous to
   the unit cell with $l$ places for electrons in an ordinary crystal without
   magnetic field. That gives the additional energy levels. But the number
   of the states for the given energy level is one per each unit cell of 
   the vortex lattice. Thus  Hamiltonian (\ref{h'}) corresponds to an "empty"
   lattice with $N^2$ states with the same energy. The spectrum of this 
   Hamiltonian has no equidistant energy levels that is valid only for the 
   oscillator problem.

   The limitation to the single magnetic cell $N{\vec \tau}_1,N{\vec \tau}_2$ 
   can be easy removed by the consideration of the vortex lattices with 
   dimensions $N_1{\vec \tau}_1,N_2{\vec \tau}_2$ where $N_1=NM_1$, $N_2=NM_2$
   for integer $M_1$,$M_2$. The representations of the larger group of
   $N_1\times N_2$ operations can be formed from the already discussed.
   
   For this
   group there are $M_1M_2$ representation of dimensionality $N$. The matrices
   corresponding to the translation ${\vec \tau_1}$,${\vec \tau_2}$ differ from
   already given only by a phase factor. These representations can be labelled
   by a vector with reciprocal space components of ${\ q}_1$,${ q}_2 $
   \begin{equation}
   \label{rp3}
   D^{\bf q}({\vec \tau_j})\equiv\exp(-i{ q_j\tau_j})D({\vec \tau_j});j=1,2
   \end{equation}
   where possible values of ${\bf q}_j$ are given by
   \begin{multline}
   \label{rp4}
   q_j=\frac{2\pi C_j}{N_j\tau_j}\\j=1,2
   \\C_1=0,1,...M_1-1;C_2=0,1,...M_2-1
   \end{multline}
   Thus each representation corresponding to a given value of ${\bf q}$ is $N$
   dimensional and one has $N$ equivalent representations in a regular one.
   The total number of the states is $M_1M_2N^2$ for the regular representation,
   also corresponds to one state per each unit cell of the vortex lattice. 
   By the construction every magnetic translation does not change the
   Hamiltonian (\ref{h'}) and therefore all states of the regular representation
   corresponds to the same energy. Therefore the spectrum is discrete and
   nonequdistant.
   If $|l|\neq 1$ where will be $|l|$ additional levels
   correswponding to $|l|$ zeros of the wave function inside the magnetic unit
   cell.

   At large magnetic fields the Hamiltonian (\ref{h'}) will be dominating in the
   full Hamiltonian (\ref{nham})
   because it linearly depends on magnetic field while the interaction term
   is proportional to the square root of it. In this case the energy of the
   ground state including the interaction can be obtained by the perturbation
   theory
   \begin{multline}
   \label{gs}
   E_0=M_1M_2N^2\epsilon_0 +\\
   \frac{1}{2}\int U_c(|{\bf r-r'}|)
   <\chi^{+}({\bf r})\chi^{+}({\bf r'})\chi({\bf r'})\chi({\bf r})>d^2rd^2r'
   \end{multline}
   here the angle brackets denote the 
   average  over the Slater determinant of the
   fully filled ground state with the energy $\epsilon_0$  of 
   the Hamiltonian 
   (\ref{h'}). The energy gap dividing the ground state from
   the next discrete level with the energy $\epsilon_1$ at large magnetic
   fields must be proportional to the value of the external magnetic field. 
   In the
   performed experiments \cite{dlg1} the linear dependence of the jump for
   electron chemical potential in strong magnetic fields  was 
   observed for the fractions 1/3 and 2/3.
   The full expression for the gap must be obtained by the numerical calculation 
   of any Bloch function for the given representation and is dependent on $K$,
   $N$,$l$ and periods $\tau_i$
   
   One can see that in the model of the vortex lattices  the gap does not depend
   exclusively on the interaction term like it was suggested in most of
   theoretical works based on the degeneracy of the ground Landau level.
   Opposite , it is almost independent from the
   interaction in strong magnetic fields. The resolution of this paradox is 
   probably the same  as in the
   rotating superfluid. The origin of the observed vortex lattices in a rotating
   superfluid is connected with the thermodynamic energy in the rotating frame
   $E'=E-{\bf \Omega M}$,where ${\bf \Omega}$ is the angular velocity
    and ${\bf M}$ is the angle momentum of the superfluid. That requires
    the superfluid velocity to be equal to the velocity of the solid body 
    rotation
    and the vortex lattice is a good approximation in a superfluid. Really it is
    connected with a different dependence of the energy on the size of the
    system giving the preference to the solid body rotation irrespective to the
    microscopical internal structure of the superfluid. 
    
    The case of magnetic 
    field differs a bit from the case of the rotation for a superfluid. The 
    quantization of 
    the orbital motion gives rise to Landau diamagnetism i.e. to the increase 
    of the system energy due to the appearance of magnetic field. It is
    possible to reduce this effect by the vortices with the opposite sign
    of  the flux.
                                                            

The previous group analysis valid for a rational number of the flux quanta
show that the energy gaps are opened at the special electron densities
corresponding to one electron per each unit cell of the vortex lattice, that
gives according to Eq.(\ref{phi'}) the electron
density 
\begin{equation}
\label{ed}
 n_e=\frac{B}{\Phi_0}\frac{N}{l-NK}
 \end{equation}
The occurrence of any
specific numbers of vortex flux quanta can be dictated  
by the ground-state energy . 
The observed fractions in FQHE correspond to the 
following tables

\vspace{5mm}
 
$K=-2,\;\;l=1$

\vspace{3mm}

\begin{tabular}{|c|c|c|c|c|c|c|c|c|c|}
\hline
$N$ & 1&2&3&-5&-2&-3&-4&4&$\infty$\\
\hline 
$\nu$
&$\frac{\mathstrut 1}{\mathstrut 3}$
&$\frac{2}{5}$
&$\frac{3}{7}$
&$\frac{5}{9}$
&$\frac{2}{3}$
&$\frac{3}{5}$
&$\frac{4}{7}$
&$\frac{4}{9}$
&$\frac{1}{2}$\\
\hline
\end{tabular}

\vspace{5mm}

That fractions correspond to celebrated Jain's rule \cite{j2}. Half filling of the 
Landau level $n_e=\frac{B}{2\phi_0}$ in the external field corresponds to a
vanishingly small effective magnetic field (zero number of flux
quanta per elementary cell).

Other observed fractions correspond to
\vspace{5mm}

$K=-1,\;\;l=1$
\vspace{3mm}

\begin{tabular}{|c|c|c|c|}
\hline
$N$&-4&4&2\\
\hline
$\nu$&$\frac{\mathstrut 4}{\mathstrut 3}$&$\frac{4}{5}$&$\frac{2}{3}$\\
\hline
\end{tabular}

\vspace{5mm}
where one has double of the fraction 2/3, and

\vspace{5mm}
$K=-1,\;\;l=2$

\vspace{3mm}
\begin{tabular}{|c|c|c|c|c|}
\hline
$N$&-7&-5&5&2\\
\hline
$\nu$&$\frac{\mathstrut 7}{\mathstrut 5}$&$\frac{5}{3}$&$\frac{5}{7}$&$\frac{1}{2}$\\
\hline
\end{tabular}

\vspace{5mm}
here one has not observed double of the fraction 1/2 with the gap
($B_{eff}\neq0$).

Thus, I have reproduced the key statement of the Jain's theory of composite
fermions  and obtained the explanation of practically all 
observed fractions
at moderate Landau levels filling in an unified frame without any hierarchial schemes.
 Of course, these results are quite crude and, in some points
hypothetical. The energy gap, the properties of elementary
charge and collective excitations, and the conductivity calculations, as well as the
analysis of different $K$ and $l,N$ values are still open questions. 
The approach to these problems needs some extensive numerical calculations. The 
preliminary results where
published in \cite{io1}. The degeneracy of the ground state in a periodic 
magnetic field was established previously for Pauli equation \cite{d}.

I am grateful to V.G. Dolgopolov, V.F. Gantmakher, V.B. Timofeev,
V.I.Marchenko and
M.V. Feigelman for helpful discussions. This work was supported by
the Russian Foundation for Basic Research ,the Program of the Presidium RAS,
and the Program of OFN RAS.
the Program for Supporting Scientific Schools.

\end{document}